# MUTUAL SUPPORT IN AGENT NETWORKS

Wan Ahmad Tajuddin Wan Abdullah and  A .K. M. Azhar*
*Dept of Physics, Universiti Malaya, 50603 Kuala Lumpur*
*\*Graduate School of Management, Universiti Putra Malaysia, 43400 Serdang*

**Abstract.** We construct a model of social behaviour through the dynamics of interacting agents. The agents undergo game-theoretic interactions where each agent can decide to lend support to particular other agents or otherwise, and agents are rewarded according to total support received. We analyse and carry out Monte Carlo simulations of such systems to uncover their evolutionary dynamics, and to explore their phase structure.

## I. INTRODUCTION

There has been increased interest recently in applying physics to the study of social and economic systems. In particular, it is hoped that statistical physics, or an extension of it, can model at least generic properties of such systems.

One approach has been the use of agent models. Here a collection of agents, whether in free space or placed on lattices or on networks, interact with each other. To encompass complex human interactions into these models, we require to implement complex interactions between agents, beyond e.g. collisions as with molecules in a gas, or magnetic interactions as in Ising spin systems and neural networks. Human interactions involve making choices, so an appropriate interaction model is that of game theory. Game-theoretic interactions would entail variables associated with strategies (choices of interaction rules), memories (internal states of agents), information (pertaining to internal state of engaged agent), and learning (optimization of strategies).

Previous studies have included prisoner's dilemma on a lattice (see e.g. the pioneering work of Axelrod[1], and elsewhere[2]), substantial work[3] on the minority game[4] and its modifications, and new constructions like the deviant's dilemma[5,6]. Here we present a new model designed to contain features of a certain aspect of society, namely rise to affluence through mutual support.

## II. THE MODEL

We have agents $i$, $i = 1,...,N$, each with a measure of influence $\phi(i)$, moving freely in free space. Random meetings of $i$ with $j$ forces $i$ to decide either to support $j$ or otherwise; this $i$ does by referring to its strategy matrix $\sigma(i,\Sigma)$, where $\Sigma$ is a situational 6-digit binary number with digits $b_5 b_4 b_3 b_2 b_1 b_0$ where $b_5$ and $b_4$ are the current values of $\pi(i,j)$ and $\pi(j,i)$, $b_3$ is 1 iff $|\phi(i)-\phi(j)| < \alpha\phi(i)$ previously, and $b_2$ is 1 iff $\phi(i) > \phi(j)$ previously, and $b_1$ and $b_0$ are corresponding flags for the current state, telling if $|\phi(i)-\phi(j)| < \alpha\phi(i)$ and if $\phi(i) > \phi(j)$ now. The support matrix $\pi(i,j)$ is also binary, where $\pi(i,j) = 1$ if $i$ supports $j$, and 0 otherwise.

When faced with a decision, $i$ would support $j$ with probability $\dfrac{0.5\nu + |\sigma(i,\Sigma)|}{\nu + |\sigma(i,\Sigma)|}$. $\nu$ is a noise parameter. Agent $j$ then has increased influence due to this support, $\phi(j) \to \phi(j) + \alpha \pi(i,j)\phi(i)$ where $\alpha$ is a constant. Agents are then ranked according to their influence $\phi(i)$, and carry out reinforcement learning quantified by their rank change, $\sigma(i,\Sigma) \to \sigma(i,\Sigma) + R_i' - R_i$.

While the essence of a game is present in the interaction in this model, it is actually more complicated as the payoffs, as in the minority game, depend on the overall action of the agents and are not static. There is also the essence of a dilemma as supporting others would jeopardise one's own standing.

## III. COMPUTATIONAL STUDIES

We carry out Monte Carlo simulations of the model to explore the system dynamics. We chose $N = 100$ and completed 200 iterations (interactions per pair) each time. To prevent blow-up, we normalize $\phi$ after each learning cycle. Learning was carried out based on the move choices 1 timestep ago so that the effect of those choices can be manifest.

Fig. 1 shows a histogram of the final distribution of influences for $\alpha = 0.3$ and $\nu = 10.0$. It shows that some structure has formed, where some agents have risen above the rest in terms of influence. In terms of strategies, it seems that certain ones emerge from the learning, as portrayed in Table 1. Winning strategies, shown in Table 2, are similar to those used by the general masses. The most prominent strategy emerging is that of continual support, which is support when $\Sigma$ = 100000. Another observation is that support in a certain situation turns to non-support in a similar situation but where the executing agent is currently more influential.

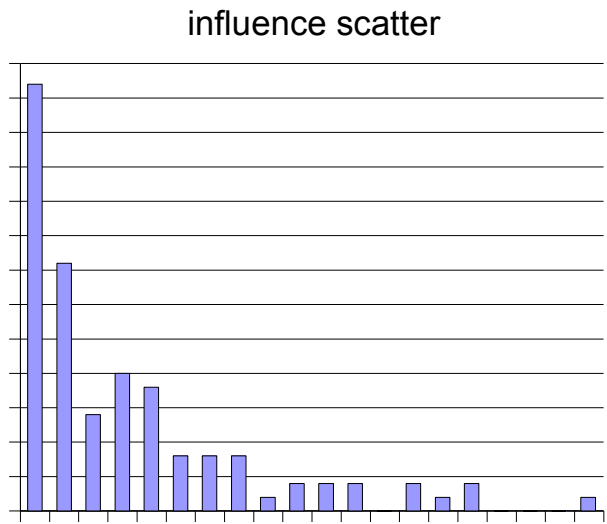

**Fig. 1** Distribution of influences

| $\Sigma$ | | | | | | | | | | no. of agents with $\sigma(i,\Sigma)$ in given range | | | | | | | | | |
|---|---|---|---|---|---|---|---|---|---|---|---|---|---|---|---|---|---|---|---|
| 000000 | 0 | 0 | 0 | 0 | 0 | 0 | 0 | 0 | 1 | 10 | 51 | 32 | 6 | 0 | 0 | 0 | 0 | 0 | 0 |
| 000001 | 0 | 0 | 0 | 0 | 0 | 0 | 0 | 0 | 2 | 79 | 19 | 0 | 0 | 0 | 0 | 0 | 0 | 0 | 0 |
| 000010 | 0 | 0 | 0 | 0 | 0 | 0 | 0 | 0 | 0 | 58 | 42 | 0 | 0 | 0 | 0 | 0 | 0 | 0 | 0 |
| 000011 | 0 | 0 | 0 | 0 | 0 | 0 | 0 | 0 | 0 | 66 | 34 | 0 | 0 | 0 | 0 | 0 | 0 | 0 | 0 |
| 000100 | 0 | 0 | 0 | 0 | 0 | 0 | 0 | 0 | 0 | 3 | 23 | 38 | 28 | 8 | 0 | 0 | 0 | 0 | 0 |
| 000101 | 0 | 0 | 0 | 0 | 0 | 0 | 3 | 33 | 50 | 13 | 1 | 0 | 0 | 0 | 0 | 0 | 0 | 0 | 0 |
| 000110 | 0 | 0 | 0 | 0 | 0 | 0 | 0 | 0 | 0 | 54 | 46 | 0 | 0 | 0 | 0 | 0 | 0 | 0 | 0 |
| 000111 | 0 | 0 | 0 | 0 | 0 | 0 | 0 | 0 | 6 | 66 | 27 | 1 | 0 | 0 | 0 | 0 | 0 | 0 | 0 |
| 001000 | 0 | 0 | 0 | 0 | 0 | 0 | 0 | 0 | 0 | 23 | 68 | 9 | 0 | 0 | 0 | 0 | 0 | 0 | 0 |
| 001001 | 0 | 0 | 0 | 0 | 0 | 0 | 0 | 0 | 0 | 76 | 24 | 0 | 0 | 0 | 0 | 0 | 0 | 0 | 0 |
| 001010 | 0 | 0 | 0 | 0 | 0 | 0 | 0 | 0 | 0 | 54 | 46 | 0 | 0 | 0 | 0 | 0 | 0 | 0 | 0 |
| 001011 | 0 | 0 | 0 | 0 | 0 | 0 | 0 | 0 | 0 | 76 | 24 | 0 | 0 | 0 | 0 | 0 | 0 | 0 | 0 |
| 001100 | 0 | 0 | 0 | 0 | 0 | 0 | 0 | 0 | 0 | 26 | 54 | 19 | 1 | 0 | 0 | 0 | 0 | 0 | 0 |
| 001101 | 0 | 0 | 0 | 0 | 0 | 0 | 0 | 0 | 9 | 73 | 18 | 0 | 0 | 0 | 0 | 0 | 0 | 0 | 0 |
| 001110 | 0 | 0 | 0 | 0 | 0 | 0 | 0 | 0 | 0 | 53 | 47 | 0 | 0 | 0 | 0 | 0 | 0 | 0 | 0 |
| 001111 | 0 | 0 | 0 | 0 | 0 | 0 | 0 | 0 | 1 | 57 | 42 | 0 | 0 | 0 | 0 | 0 | 0 | 0 | 0 |
| 010000 | 0 | 0 | 0 | 0 | 0 | 0 | 0 | 0 | 0 | 11 | 72 | 16 | 1 | 0 | 0 | 0 | 0 | 0 | 0 |
| 010001 | 0 | 0 | 0 | 0 | 0 | 0 | 0 | 0 | 3 | 77 | 20 | 0 | 0 | 0 | 0 | 0 | 0 | 0 | 0 |
| 010010 | 0 | 0 | 0 | 0 | 0 | 0 | 0 | 0 | 0 | 58 | 42 | 0 | 0 | 0 | 0 | 0 | 0 | 0 | 0 |
| 010011 | 0 | 0 | 0 | 0 | 0 | 0 | 0 | 0 | 0 | 56 | 44 | 0 | 0 | 0 | 0 | 0 | 0 | 0 | 0 |
| 010100 | 0 | 0 | 0 | 0 | 0 | 0 | 0 | 0 | 0 | 1 | 23 | 38 | 31 | 6 | 1 | 0 | 0 | 0 | 0 |
| 010101 | 0 | 0 | 0 | 2 | 9 | 25 | 38 | 14 | 10 | 2 | 0 | 0 | 0 | 0 | 0 | 0 | 0 | 0 | 0 |
| 010110 | 0 | 0 | 0 | 0 | 0 | 0 | 0 | 0 | 1 | 51 | 47 | 1 | 0 | 0 | 0 | 0 | 0 | 0 | 0 |
| 010111 | 0 | 0 | 0 | 0 | 0 | 0 | 0 | 0 | 19 | 53 | 25 | 3 | 0 | 0 | 0 | 0 | 0 | 0 | 0 |
| 011000 | 0 | 0 | 0 | 0 | 0 | 0 | 0 | 0 | 0 | 36 | 58 | 6 | 0 | 0 | 0 | 0 | 0 | 0 | 0 |
| 011001 | 0 | 0 | 0 | 0 | 0 | 0 | 0 | 0 | 1 | 82 | 17 | 0 | 0 | 0 | 0 | 0 | 0 | 0 | 0 |
| 011010 | 0 | 0 | 0 | 0 | 0 | 0 | 0 | 0 | 0 | 58 | 42 | 0 | 0 | 0 | 0 | 0 | 0 | 0 | 0 |
| 011011 | 0 | 0 | 0 | 0 | 0 | 0 | 0 | 0 | 0 | 64 | 36 | 0 | 0 | 0 | 0 | 0 | 0 | 0 | 0 |
| 011100 | 0 | 0 | 0 | 0 | 0 | 0 | 0 | 0 | 0 | 15 | 66 | 19 | 0 | 0 | 0 | 0 | 0 | 0 | 0 |

| Σ | | | | | | | | | | | | | | | | | | | |
|---|---|---|---|---|---|---|---|---|---|---|---|---|---|---|---|---|---|---|---|
| 011101 | 0 | 0 | 0 | 0 | 0 | 0 | 0 | 0 | 13 | 73 | 14 | 0 | 0 | 0 | 0 | 0 | 0 | 0 | 0 |
| 011110 | 0 | 0 | 0 | 0 | 0 | 0 | 0 | 0 | 0 | 50 | 50 | 0 | 0 | 0 | 0 | 0 | 0 | 0 | 0 |
| 011111 | 0 | 0 | 0 | 0 | 0 | 0 | 0 | 0 | 0 | 62 | 38 | 0 | 0 | 0 | 0 | 0 | 0 | 0 | 0 |
| 100000 | 0 | 0 | 0 | 0 | 0 | 0 | 0 | 0 | 0 | 1 | 1 | 3 | 11 | 12 | 29 | 18 | 13 | 10 | 1 | 1 |
| 100001 | 0 | 0 | 0 | 0 | 0 | 0 | 1 | 19 | 56 | 23 | 1 | 0 | 0 | 0 | 0 | 0 | 0 | 0 | 0 |
| 100010 | 0 | 0 | 0 | 0 | 0 | 0 | 0 | 0 | 3 | 55 | 41 | 1 | 0 | 0 | 0 | 0 | 0 | 0 | 0 |
| 100011 | 0 | 0 | 0 | 0 | 0 | 0 | 0 | 0 | 21 | 59 | 17 | 3 | 0 | 0 | 0 | 0 | 0 | 0 | 0 |
| 100100 | 0 | 0 | 0 | 0 | 0 | 0 | 0 | 0 | 0 | 19 | 70 | 11 | 0 | 0 | 0 | 0 | 0 | 0 | 0 |
| 100101 | 0 | 0 | 0 | 0 | 0 | 0 | 0 | 0 | 5 | 80 | 15 | 0 | 0 | 0 | 0 | 0 | 0 | 0 | 0 |
| 100110 | 0 | 0 | 0 | 0 | 0 | 0 | 0 | 0 | 0 | 62 | 38 | 0 | 0 | 0 | 0 | 0 | 0 | 0 | 0 |
| 100111 | 0 | 0 | 0 | 0 | 0 | 0 | 0 | 0 | 0 | 58 | 42 | 0 | 0 | 0 | 0 | 0 | 0 | 0 | 0 |
| 101000 | 0 | 0 | 0 | 0 | 0 | 0 | 0 | 0 | 0 | 18 | 70 | 12 | 0 | 0 | 0 | 0 | 0 | 0 | 0 |
| 101001 | 0 | 0 | 0 | 0 | 0 | 0 | 0 | 0 | 0 | 77 | 23 | 0 | 0 | 0 | 0 | 0 | 0 | 0 | 0 |
| 101010 | 0 | 0 | 0 | 0 | 0 | 0 | 0 | 0 | 0 | 52 | 48 | 0 | 0 | 0 | 0 | 0 | 0 | 0 | 0 |
| 101011 | 0 | 0 | 0 | 0 | 0 | 0 | 0 | 0 | 0 | 62 | 38 | 0 | 0 | 0 | 0 | 0 | 0 | 0 | 0 |
| 101100 | 0 | 0 | 0 | 0 | 0 | 0 | 0 | 0 | 0 | 16 | 76 | 8 | 0 | 0 | 0 | 0 | 0 | 0 | 0 |
| 101101 | 0 | 0 | 0 | 0 | 0 | 0 | 0 | 0 | 0 | 79 | 21 | 0 | 0 | 0 | 0 | 0 | 0 | 0 | 0 |
| 101110 | 0 | 0 | 0 | 0 | 0 | 0 | 0 | 0 | 0 | 54 | 46 | 0 | 0 | 0 | 0 | 0 | 0 | 0 | 0 |
| 101111 | 0 | 0 | 0 | 0 | 0 | 0 | 0 | 0 | 0 | 63 | 37 | 0 | 0 | 0 | 0 | 0 | 0 | 0 | 0 |
| 110000 | 0 | 0 | 0 | 0 | 0 | 0 | 0 | 0 | 0 | 1 | 13 | 38 | 26 | 17 | 5 | 0 | 0 | 0 | 0 |
| 110001 | 0 | 0 | 0 | 0 | 0 | 0 | 0 | 20 | 45 | 29 | 6 | 0 | 0 | 0 | 0 | 0 | 0 | 0 | 0 |
| 110010 | 0 | 0 | 0 | 0 | 0 | 0 | 0 | 0 | 0 | 65 | 34 | 1 | 0 | 0 | 0 | 0 | 0 | 0 | 0 |
| 110011 | 0 | 0 | 0 | 0 | 0 | 0 | 0 | 0 | 15 | 58 | 25 | 2 | 0 | 0 | 0 | 0 | 0 | 0 | 0 |
| 110100 | 0 | 0 | 0 | 0 | 0 | 0 | 0 | 0 | 0 | 21 | 73 | 6 | 0 | 0 | 0 | 0 | 0 | 0 | 0 |
| 110101 | 0 | 0 | 0 | 0 | 0 | 0 | 0 | 0 | 15 | 78 | 7 | 0 | 0 | 0 | 0 | 0 | 0 | 0 | 0 |
| 110110 | 0 | 0 | 0 | 0 | 0 | 0 | 0 | 0 | 0 | 49 | 51 | 0 | 0 | 0 | 0 | 0 | 0 | 0 | 0 |
| 110111 | 0 | 0 | 0 | 0 | 0 | 0 | 0 | 0 | 1 | 65 | 34 | 0 | 0 | 0 | 0 | 0 | 0 | 0 | 0 |
| 111000 | 0 | 0 | 0 | 0 | 0 | 0 | 0 | 0 | 0 | 42 | 54 | 4 | 0 | 0 | 0 | 0 | 0 | 0 | 0 |
| 111001 | 0 | 0 | 0 | 0 | 0 | 0 | 0 | 0 | 0 | 74 | 26 | 0 | 0 | 0 | 0 | 0 | 0 | 0 | 0 |
| 111010 | 0 | 0 | 0 | 0 | 0 | 0 | 0 | 0 | 0 | 52 | 48 | 0 | 0 | 0 | 0 | 0 | 0 | 0 | 0 |
| 111011 | 0 | 0 | 0 | 0 | 0 | 0 | 0 | 0 | 0 | 63 | 37 | 0 | 0 | 0 | 0 | 0 | 0 | 0 | 0 |
| 111100 | 0 | 0 | 0 | 0 | 0 | 0 | 0 | 0 | 0 | 25 | 68 | 7 | 0 | 0 | 0 | 0 | 0 | 0 | 0 |
| 111101 | 0 | 0 | 0 | 0 | 0 | 0 | 0 | 0 | 3 | 78 | 19 | 0 | 0 | 0 | 0 | 0 | 0 | 0 | 0 |
| 111110 | 0 | 0 | 0 | 0 | 0 | 0 | 0 | 0 | 0 | 66 | 34 | 0 | 0 | 0 | 0 | 0 | 0 | 0 | 0 |
| 111111 | 0 | 0 | 0 | 0 | 0 | 0 | 0 | 0 | 0 | 77 | 22 | 1 | 0 | 0 | 0 | 0 | 0 | 0 | 0 |

**Table 1** Final strategy distribution for all agents. The 2 middle bins are within the noise level from 0.

| Σ | no. of agents with $\sigma(i,\Sigma)$ in given range | | | | | | | | | | | | | | | | | | |
|---|---|---|---|---|---|---|---|---|---|---|---|---|---|---|---|---|---|---|---|
| 000000 | 0 | 0 | 0 | 0 | 0 | 0 | 0 | 0 | 0 | 3 | 4 | 3 | 0 | 0 | 0 | 0 | 0 | 0 | 0 |
| 000001 | 0 | 0 | 0 | 0 | 0 | 0 | 0 | 0 | 0 | 9 | 1 | 0 | 0 | 0 | 0 | 0 | 0 | 0 | 0 |
| 000010 | 0 | 0 | 0 | 0 | 0 | 0 | 0 | 0 | 0 | 7 | 3 | 0 | 0 | 0 | 0 | 0 | 0 | 0 | 0 |
| 000011 | 0 | 0 | 0 | 0 | 0 | 0 | 0 | 0 | 0 | 7 | 3 | 0 | 0 | 0 | 0 | 0 | 0 | 0 | 0 |
| 000100 | 0 | 0 | 0 | 0 | 0 | 0 | 0 | 0 | 0 | 0 | 3 | 3 | 3 | 1 | 0 | 0 | 0 | 0 | 0 |
| 000101 | 0 | 0 | 0 | 0 | 0 | 0 | 1 | 0 | 6 | 3 | 0 | 0 | 0 | 0 | 0 | 0 | 0 | 0 | 0 |
| 000110 | 0 | 0 | 0 | 0 | 0 | 0 | 0 | 0 | 0 | 5 | 5 | 0 | 0 | 0 | 0 | 0 | 0 | 0 | 0 |
| 000111 | 0 | 0 | 0 | 0 | 0 | 0 | 0 | 0 | 1 | 6 | 3 | 0 | 0 | 0 | 0 | 0 | 0 | 0 | 0 |
| 001000 | 0 | 0 | 0 | 0 | 0 | 0 | 0 | 0 | 0 | 2 | 7 | 1 | 0 | 0 | 0 | 0 | 0 | 0 | 0 |
| 001001 | 0 | 0 | 0 | 0 | 0 | 0 | 0 | 0 | 0 | 8 | 2 | 0 | 0 | 0 | 0 | 0 | 0 | 0 | 0 |
| 001010 | 0 | 0 | 0 | 0 | 0 | 0 | 0 | 0 | 0 | 4 | 6 | 0 | 0 | 0 | 0 | 0 | 0 | 0 | 0 |
| 001011 | 0 | 0 | 0 | 0 | 0 | 0 | 0 | 0 | 0 | 9 | 1 | 0 | 0 | 0 | 0 | 0 | 0 | 0 | 0 |
| 001100 | 0 | 0 | 0 | 0 | 0 | 0 | 0 | 0 | 0 | 1 | 6 | 2 | 1 | 0 | 0 | 0 | 0 | 0 | 0 |
| 001101 | 0 | 0 | 0 | 0 | 0 | 0 | 0 | 0 | 1 | 8 | 1 | 0 | 0 | 0 | 0 | 0 | 0 | 0 | 0 |
| 001110 | 0 | 0 | 0 | 0 | 0 | 0 | 0 | 0 | 0 | 7 | 3 | 0 | 0 | 0 | 0 | 0 | 0 | 0 | 0 |
| 001111 | 0 | 0 | 0 | 0 | 0 | 0 | 0 | 0 | 1 | 3 | 6 | 0 | 0 | 0 | 0 | 0 | 0 | 0 | 0 |
| 010000 | 0 | 0 | 0 | 0 | 0 | 0 | 0 | 0 | 0 | 1 | 8 | 1 | 0 | 0 | 0 | 0 | 0 | 0 | 0 |
| 010001 | 0 | 0 | 0 | 0 | 0 | 0 | 0 | 0 | 0 | 8 | 2 | 0 | 0 | 0 | 0 | 0 | 0 | 0 | 0 |
| 010010 | 0 | 0 | 0 | 0 | 0 | 0 | 0 | 0 | 0 | 5 | 5 | 0 | 0 | 0 | 0 | 0 | 0 | 0 | 0 |
| 010011 | 0 | 0 | 0 | 0 | 0 | 0 | 0 | 0 | 0 | 7 | 3 | 0 | 0 | 0 | 0 | 0 | 0 | 0 | 0 |
| 010100 | 0 | 0 | 0 | 0 | 0 | 0 | 0 | 0 | 0 | 0 | 2 | 5 | 1 | 1 | 1 | 0 | 0 | 0 | 0 |
| 010101 | 0 | 0 | 0 | 0 | 2 | 3 | 2 | 2 | 1 | 0 | 0 | 0 | 0 | 0 | 0 | 0 | 0 | 0 | 0 |
| 010110 | 0 | 0 | 0 | 0 | 0 | 0 | 0 | 0 | 1 | 4 | 5 | 0 | 0 | 0 | 0 | 0 | 0 | 0 | 0 |
| 010111 | 0 | 0 | 0 | 0 | 0 | 0 | 0 | 0 | 2 | 7 | 1 | 0 | 0 | 0 | 0 | 0 | 0 | 0 | 0 |
| 011000 | 0 | 0 | 0 | 0 | 0 | 0 | 0 | 0 | 0 | 2 | 8 | 0 | 0 | 0 | 0 | 0 | 0 | 0 | 0 |
| 011001 | 0 | 0 | 0 | 0 | 0 | 0 | 0 | 0 | 0 | 8 | 2 | 0 | 0 | 0 | 0 | 0 | 0 | 0 | 0 |

```
011010   0  0  0  0  0  0  0  0  0   9  1  0  0  0  0  0  0  0  0
011011   0  0  0  0  0  0  0  0  0   7  3  0  0  0  0  0  0  0  0
011100   0  0  0  0  0  0  0  0  0   1  9  0  0  0  0  0  0  0  0
011101   0  0  0  0  0  0  0  0  0  10  0  0  0  0  0  0  0  0  0
011110   0  0  0  0  0  0  0  0  0   5  5  0  0  0  0  0  0  0  0
011111   0  0  0  0  0  0  0  0  0   7  3  0  0  0  0  0  0  0  0
100000   0  0  0  0  0  0  0  0  0   0  0  0  1  3  3  2  0  0  1
100001   0  0  0  0  0  0  0  1  7   2  0  0  0  0  0  0  0  0  0
100010   0  0  0  0  0  0  0  0  0   6  4  0  0  0  0  0  0  0  0
100011   0  0  0  0  0  0  0  0  3   6  1  0  0  0  0  0  0  0  0
100100   0  0  0  0  0  0  0  0  0   2  8  0  0  0  0  0  0  0  0
100101   0  0  0  0  0  0  0  0  0   8  2  0  0  0  0  0  0  0  0
100110   0  0  0  0  0  0  0  0  0   5  5  0  0  0  0  0  0  0  0
100111   0  0  0  0  0  0  0  0  0   4  6  0  0  0  0  0  0  0  0
101000   0  0  0  0  0  0  0  0  0   4  5  1  0  0  0  0  0  0  0
101001   0  0  0  0  0  0  0  0  0   7  3  0  0  0  0  0  0  0  0
101010   0  0  0  0  0  0  0  0  0   6  4  0  0  0  0  0  0  0  0
101011   0  0  0  0  0  0  0  0  0   5  5  0  0  0  0  0  0  0  0
101100   0  0  0  0  0  0  0  0  0   0  9  1  0  0  0  0  0  0  0
101101   0  0  0  0  0  0  0  0  0   9  1  0  0  0  0  0  0  0  0
101110   0  0  0  0  0  0  0  0  0   3  7  0  0  0  0  0  0  0  0
101111   0  0  0  0  0  0  0  0  0   7  3  0  0  0  0  0  0  0  0
110000   0  0  0  0  0  0  0  0  0   2  4  3  1  0  0  0  0  0  0
110001   0  0  0  0  0  0  0  2  5   2  1  0  0  0  0  0  0  0  0
110010   0  0  0  0  0  0  0  0  0   5  5  0  0  0  0  0  0  0  0
110011   0  0  0  0  0  0  0  0  0   6  3  1  0  0  0  0  0  0  0
110100   0  0  0  0  0  0  0  0  0   2  8  0  0  0  0  0  0  0  0
110101   0  0  0  0  0  0  0  0  1   8  1  0  0  0  0  0  0  0  0
110110   0  0  0  0  0  0  0  0  0   5  5  0  0  0  0  0  0  0  0
110111   0  0  0  0  0  0  0  0  0   7  3  0  0  0  0  0  0  0  0
111000   0  0  0  0  0  0  0  0  0   6  2  2  0  0  0  0  0  0  0
111001   0  0  0  0  0  0  0  0  0   7  3  0  0  0  0  0  0  0  0
111010   0  0  0  0  0  0  0  0  0   4  6  0  0  0  0  0  0  0  0
111011   0  0  0  0  0  0  0  0  0   6  4  0  0  0  0  0  0  0  0
111100   0  0  0  0  0  0  0  0  0   3  7  0  0  0  0  0  0  0  0
111101   0  0  0  0  0  0  0  0  0  10  0  0  0  0  0  0  0  0  0
111110   0  0  0  0  0  0  0  0  0   5  5  0  0  0  0  0  0  0  0
111111   0  0  0  0  0  0  0  0  0   7  3  0  0  0  0  0  0  0  0
```

**Table 2** Final strategy distribution for agents in the top 10% ranking. The 2 middle bins are within the noise level from 0.

To examine how support evolves, we plot Fig. 2 which shows the average magnitude of support, as well as that of its asymmetry. The asymmetry tells us whether mutual support (low asymmetry) or the opposite occurs. It seems that on average, the absolute amount of support remains constant, indicating that the society does not degenerate into non-interacting individuals, but the asymmetry increases with time, indicating that one-way support increases in prominence.

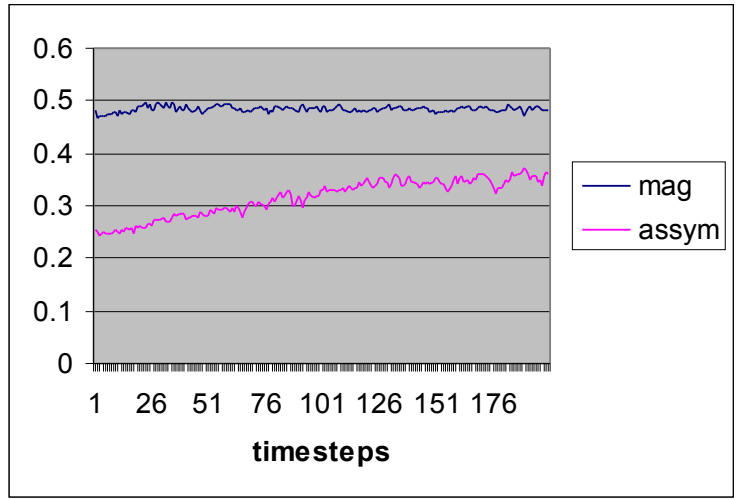

**Fig. 2** Support evolution

For differing parameters, the behaviour displayed by the system does not seem to change substantially. We find similar results with learning based on immediate move, learning based on move 2 timesteps before, with $\alpha = 0.1$, $\alpha = 0.5$, and with $v = 5.0$ and $v = 20.0$.

## IV. CONCLUSIONS

We have studied an agent model embodying mutual support and observe the emergence of structures or heterogeneities in the influence distribution. More prominent was the emergence of successful (?) strategies. So far, the phase space of the system seem homogeneous.

The emergence of strategies is interesting and it would be nice to understand the rationale for strategies, and then to relate to real social dynamics.